\documentclass[prx,preprintnumbers,
twocolumn,
eqsecnum,floatfix,letterpaper,superscriptaddress,nofootinbib]{revtex4}
\usepackage{amsmath,amssymb,graphicx}
\usepackage[printwatermark]{xwatermark}
\usepackage{xcolor}
\usepackage{lipsum}
\usepackage{bm}
\usepackage{multirow}
\usepackage{color}
\usepackage{booktabs}
\usepackage{tabularx}
\usepackage{times}
\usepackage{microtype}
\usepackage{booktabs}
\usepackage{subfigure}
\usepackage{csquotes}
\usepackage[normalem]{ulem}
\usepackage[varg]{txfonts}
\usepackage{dirtytalk}
\usepackage[colorlinks, pdfborder={0 0 0}]{hyperref}
\definecolor{LinkColor}{rgb}{0.75 , 0, 0}
\definecolor{CiteColor}{rgb}{0, 0.5, 0.5}
\definecolor{UrlColor}{rgb}{0, 0, 0.75}
\hypersetup{linkcolor=LinkColor}
\hypersetup{citecolor=CiteColor}
\hypersetup{urlcolor=UrlColor}

\allowdisplaybreaks
\newcommand{\FigStart}{\begin{figure*}[h]}
\newcommand{\FigEnd}{\end{figure*}}

\newcommand{\msun}{$M_{\odot}$} 
\newcommand{\Msun}{M_{\odot}} 
\newcommand{\EnumStart}{\begin{enumerate}}
\newcommand{\EnumEnd}{\end{enumerate}}

\newcommand{\Mc}{\mathcal{M}}
\newcommand{\fupper}{f_\text{upper}}
\newcommand{\flower}{f_\text{lower}}

\def\ks{\kappa_s}

\def\dk{\delta\kappa}
\def\dks{\delta\kappa_s}

\def\dk1{\delta\kappa_1}
\def\dk2{\delta\kappa_2}

\def\effspin{\chi_{\text{eff}}}

\def\lalinference{\texttt{LALInference }}
\def\imrPhenomPv2{\texttt{IMRPhenomPv2}}
\def\taylorF2{\texttt{TaylorF2}}

\def\Hyp{\mathcal{H}}

\def\Z{\mathcal{Z}}

\def\Hgeneric{\mathcal{H}_{\text{\tiny NBH}}}
\def\Hgr{\mathcal{H}_{\text{\tiny BH}}}

\def\paramVec{\bar{\theta}}

\def\BnonGRvsGR{\mathcal{B}^{\text{\tiny NBH}}_{\text{\tiny BH}}}
\def\BgenericVsGR{\mathcal{B}^{\text{\tiny NBH}}_{\text{\tiny BH}}}

\def\Zalt{\mathcal{Z}_{\text{\tiny NBH}}}
\def\Zgr{\mathcal{Z}_{\text{\tiny BH}}}

\def\thetaBH{\paramVec_{\text{\tiny BH}}}  
\def\thetaNBH{\paramVec_{\text{\tiny NBH}}}  
\def\hBH{\tilde{h}^{\text{\tiny BH}}}
\def\hNBH{\tilde{h}^{\text{\tiny NBH}}}
\def\ff{\rm{FF}} 
\def\ffNBH{\ff(\thetaNBH)}

\def\th{\tilde{{h}}}

\def\td{\tilde{{d}}}

\def\Mpc{\text{~Mpc}}

\def\snf{{S_{n}}(f)}

\def\wrt{\textit{w.r.t }}
\def\thatis{\textit{i.e., }}
\def\etc{\textit{etc. }}

\begin{document}
\title{Constraints on the binary black hole nature of GW151226 and
GW170608 from the measurement of spin-induced quadrupole moments} 
\author{N. V. Krishnendu}
\affiliation{Chennai Mathematical Institute, Siruseri 603103, Tamilnadu,
India}
\author{M. Saleem}
\affiliation{Chennai Mathematical Institute, Siruseri 603103, Tamilnadu,
India}
\author{A. Samajdar}
\affiliation{Nikhef, 105 Science Park, 1098 XG Amsterdam, The Netherlands}
\author{K. G. Arun}
\affiliation{Chennai Mathematical Institute, Siruseri 603103, Tamilnadu,
India}
\author{W. Del Pozzo}
\affiliation{Dipartimento di Fisica “Enrico Fermi”, Universita` di Pisa, and INFN sezione di Pisa, Pisa I-56127, Italy}
\affiliation{Institute of Gravitational Wave Astronomy, University of Birmingham, Edgbaston, Birmingham B15 2TT, United Kingdom}
\author{Chandra Kant Mishra}
\affiliation{Indian Institute of Technology Madras, Chennai, 600036, India}
\date{\today}

\begin{abstract}
	According to the ``no-hair'' conjecture, a Kerr black hole
	(BH) is completely described by its mass and spin.   In particular, the spin-induced quadrupole moment of a Kerr BH with mass $m$ and dimensionless spin $\chi$ can be written as $Q=-\kappa\,m^3\chi^2$, where $\kappa_{\rm BH}=1$. Thus by measuring the spin-induced quadrupole parameter     $\kappa$, we can test the binary black hole nature of compact binaries and distinguish them from binaries comprised of other exotic compact objects, as proposed in~\cite{Krishnendu:2017shb}.  
	Here, we present a Bayesian framework to carry out this test where we measure the symmetric combination of individual spin-induced quadrupole moment parameters fixing the anti-symmetric combination to be zero. The analysis is restricted to the inspiral part of the signal as the spin-induced deformations are not modeled in the post-inspiral regime. We perform detailed simulations to investigate the applicability of this method for compact binaries of different masses and spins and also explore various degeneracies in the parameter space which can affect this test.    We then apply this method to the gravitational wave events, GW151226 and GW170608 detected during the first and second observing runs of Advanced LIGO and Advanced Virgo detectors. We find the two events to be consistent with binary black hole mergers in general relativity. By combining information from several more of such events in future, this method can be used to set constraints on the black hole nature of the population of compact binaries that are detected by the Advanced LIGO and Advanced Virgo detectors.
\end{abstract}

 \maketitle
\section{Introduction}\label{intro}

The existence of black holes (BHs) is one of the fundamental predictions of Einstein's general theory of relativity. A classical black hole is defined by an event horizon which covers a space-time singularity. The horizon acts as a one-way membrane through which things can fall in, but nothing can come out. Recent detections of gravitational waves from binary black holes~\cite{Discovery,
	GW151226,GW170104,GW170608,GW170814,O1O2catalogLSC2018} and the radio image of the
super-massive black hole residing at the center of the galaxy M87 are
consistent with the predictions of astrophysical black holes~\cite{EHTpaper1_2019ApJ}.
Developing methods to accurately constrain the black hole
nature of compact objects is of great
importance from a fundamental physics
viewpoint (see Ref.~\cite{CardosoLivingReview} for a
review). Ruling out the Kerr nature of a black hole candidate could
hint at new and exotic physics~\cite{Giudice}. 

Parametric models describing BH mimickers (astrophysical objects which can mimic the properties of
BHs) play an important role in interpreting astrophysical observations
and permits setting constraints on the parameter space of BH mimickers.
The parametrization may be at the level of the
metric~\cite{Ryan:1995wh,Collins:2004ex,Vigeland:2009pr,Glampedakis:2005cf}
in the case of super-massive objects at the center of galaxies whereas
BH signatures in the gravitational waveforms would be best-suited for
tests of BH nature using gravitational waves. In this paper, we focus on
tests of binary black hole nature using gravitational wave observations
of compact binaries.

The Advanced LIGO and Advanced Virgo detectors have so far detected
{eleven
	gravitational wave (GW) signals from compact binary coalescences
	~\cite{Discovery, GW151226,GW170104,GW170608,GW170814,O1O2catalogLSC2018}.
	Various consistency tests have been performed on these signals
	\cite{Meidam:2014jpa,Yunes:2013dva, AIQS06a,AIQS06b,Berti:2018cxi,Berti:2018vdi,
		Will:1977wq,TIGER2014,Yunes:2009ke,Will:1997bb,Samajdar:2017mka,Ghosh:2017gfp,TOG,GW170104,GW170608,GW170814,O2TGR} which
	show that ten of them are consistent with the signals expected from
	binary black holes whereas one of them is consistent with binary
	neutron star coalescence.  However, the uncertainties on
possible deviations from general relativity reported in these works are large enough
that black hole mimicker models such as boson stars~\cite{Ryan97b} or
gravastars~\cite{gravastars,Uchikata:2015yma,Uchikata2016qku} can
fit in with some tuning of the free parameters describing these models.
Hence a dedicated analysis looking for generic signatures of black
hole mimickers is necessary to make precise quantitative statements about the
nature of the compact binaries which forms the theme for this paper.

	The ability of the gravitational wave detectors to accurately measure
	the phase of the signal can be used as a powerful tool to develop
	theoretical waveform models which parametrize different classes of BH
	mimickers.
	Different types of parametrizations have been put forward in the literature to distinguish binary black hole signals from binary black hole mimicker signals where at least one of the components of the binary is not a BH (henceforth, we also use `non-BH signals' interchangeably for BH mimicker signals). 
	All of them rely on measuring various physical effects which
	appear in the waveform and are unique signatures of BHs.
	For example,  the tidal deformability parameter of the compact binary constituents which is predicted to be zero for Kerr BHs up to next-to-next-to-leading order in the BH spin ~\cite{RelativistictheoryofTLNs_EricBinnington,NHTBHsAstrophysicalEnvironments2015,DamourNagarTLNsofNSs,Hinderer:2007mb,FlanaganHindererNSTLNs,Pani:2015hfa} while a non-zero value is expected if there are some exotic physics at play~\cite{Cardoso:2017cfl,Sennett:2017etc,Jimenez-Forteza:2018buh,Abdelsalhin:2018reg,Johnson-McDaniel:2018uvs,Porto:2016zng}. Another one is the quasi-normal mode ringdown spectrum analysis \cite{VishuNature,Berti:2005ys,Dreyer:2003bv,GossanSathya2011} of the compact object formed by the merger, which for a Kerr BH, will be uniquely determined by its mass and spin, while for non-BH objects, there will be additional dependencies~\cite{Chirenti:2007mk,BertiBS,Macedo:2013jja,Macedo:2016wgh,BertiBS,GossanVeitchSathya2011ha}. Various other methods include measuring the late ringdown modes \cite{Cardoso:2016rao,Cardoso:2016oxy}, measuring the tidal heating parameter~\cite{Hartle:1973zz,Maselli:2017cmm,Chatziioannou:2016kem,Chatziioannou:2012gq}, testing the consistency between the binary black hole parameters independently measured from the inspiral and post-merger signals~\cite{Ghosh:2017gfp}, and testing the consistency between different spherical harmonic modes using gravitational waveform models having higher modes~\cite{Dhanpal:2018ufk}. A recent proposal~\cite{Krishnendu:2017shb} to use the measurement of spin-induced multipole moments of the binary constituents as a probe of their black hole nature forms the subject of this paper.

	\subsection{Measurement of spin-induced multipole moments as a test of black hole nature}
	\label{sec-KAM-summary}
	
	Recently, Krishnendu \textit{et al.} 2017
	\cite{Krishnendu:2017shb} proposed a new method to distinguish between
	binary black holes  and binary black hole mimickers by measuring the
	spin-induced multipole moments of the compact objects. As the name
	indicates, these multipole moments arise due to the spins of the compact
	objects and their values will depend on the types of objects. The
	leading order effect is due to the spin-induced quadrupole moment which
	appears as part of the spin-spin interactions in the post-Newtonian
	phasing formula (at the second post-Newtonian [2PN] order) and is schematically represented by \cite{Poisson:1997ha}, 
	\begin{equation} 
	Q=-\kappa\,\mathbf{\chi}^2 \,m^3,
	\label{eq-leadingorder} 
	\end{equation} 
	where $Q$ is known as the spin-induced quadrupole moment scalar,
	$m$ is the mass and $\chi$ is the dimensionless spin  parameter (defined
	as $\vec{\chi}=\vec{S}/m^{2}$, where $\vec{S}$ is the spin angular
	momentum of the compact object) and $\kappa$ is the spin-induced
	quadrupole moment coefficient which measures the deformations due to the spinning motion of the object. The value of $\kappa$ for Kerr black
	holes is unique and equals unity, according to the ``no-hair" conjecture
	\cite{Hansen74,Carter71,Gurlebeck:2015xpa} while for other (non-BH)
	compact objects, it varies depending upon their internal structure. 
  For example,  $\kappa$ may vary between $\sim 2-14$ for neutron stars up to quadratic in spin, depending on the various equation of states
	~\cite{Laarakkers:1997hb,Pappas:2012qg,Pappas:2012ns}. A detailed study of the estimation of equation of state of spinning
binary neutron stars making use of the spin-induced quadrupole moment
effect can be found in~\cite{Harry:2018hke}. For slowly
	spinning boson stars, the value of $\kappa$ can roughly vary between 10
	to 150 \cite{Ryan97b} while for  gravastars~\cite{gravastars} $\kappa$
	can be negative as well (see the references
	~\cite{Uchikata:2015yma,Uchikata2016qku} for more details).  In general,
	$\kappa>0$ refers to those classes of compact objects which undergo
	\textit{oblate} deformation due to spins whereas $\kappa<0$ refers to objects whose
	spin-induced deformation is \textit{prolate} in nature. Hence, estimated upper and
	lower bounds on the value of $\kappa$ from GW observations can lead to
	constraints on the allowed parameter space of various classes of non-black hole compact objects or BH mimickers.
	
	A Fisher matrix study carried out in Krishnendu \textit{et al.}
\cite{Krishnendu:2017shb} explored the accuracy with which the
spin-induced quadrupole moment parameters can be measured for
non-precessing binaries with various masses and spins. This test was
performed with a post-Newtonian waveform  with 4PN ({\it{partial}})
phase corrections and 2PN amplitude corrections and considered a
one-parameter deformation of the binary black hole waveforms
parametrized by the symmetric combination defined by
$\kappa_s=\frac{1}{2}\left(\kappa_1+\kappa_2\right)$, where
$\kappa_{1,2}$ denote the spin-induced quadrupole moment parameters of
the binary constituents.   The study showed that with the second
generation (2G) ground-based detectors, $\kappa_s$ can be bounded to
values of the order of unity if the black holes are nearly of equal
masses and have their spins aligned with respect to the orbital angular
momentum with dimensionless magnitudes $\sim 0.9$.
In \cite{Krishnendu:2018nqa}, the authors extended this study to
demonstrate the capabilities of third-generation (3G) gravitational wave
detectors such as Einstein telescope~\cite{Sathyaprakash:2011bh} and
Cosmic Explorer \cite{Regimbau:2012ir, Hild:2010id,
Hild:2008ng,Sathyaprakash:2011bh}, to test the binary black hole nature
by measuring the spin-induced \textit{multipole} moment parameters. It
was found that the improved sensitivities of third-generation detectors
improve the overall measurement errors on spin-induced quadrupole moment
parameter compared to 2G detectors.  The symmetric combination of
the parameter can be measured to accuracy of the order of unity even if
the dimensional spins are $\sim 0.2$. Further, 3G detectors would also
allow us to simultaneously constrain both spin-induced quadrupole and
octupole moment parameters of the binary and hence constrain the
first four multipoles (mass, spin, quadrupole and octupole moments) of
the system. Further, in certain regions of binary black
hole parameter space, it gives the ability to measure the spin-induced
quadrupole moment parameters of the individual constituents of the
binary, rather than measuring the symmetric combination defined above.
More recently this study was further extended to the case of space-based
detectors LISA and DECIGO~\cite{KY19} and it was found that they offer
unprecedented opportunity to test the black hole nature of compact
binaries in the intermediate-mass and super-massive mass regimes  by
measuring the symmetric combination of the quadrupole parameter to
accuracy of the order of $0.1$ even for spin magnitudes of $\sim 0.5$.
	
	\begin{figure*}[htp] \centering
		\includegraphics[scale=0.62]{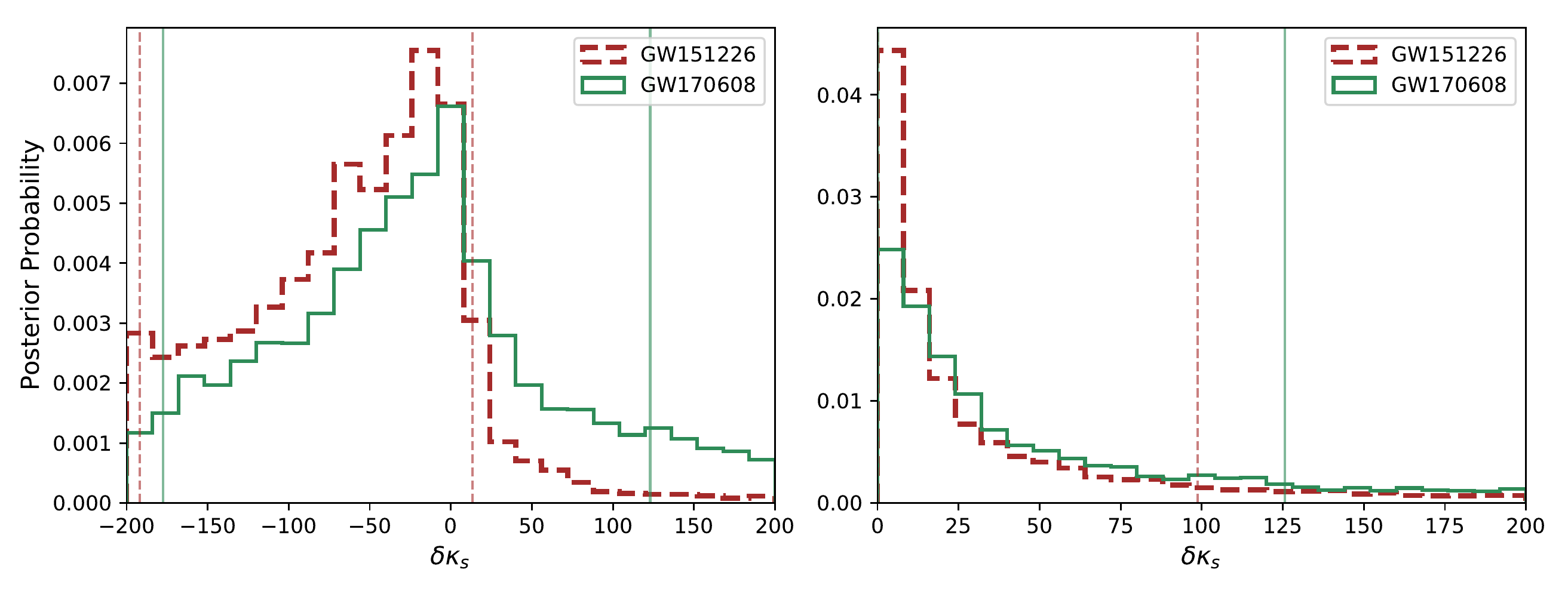} 
		\caption{ Posterior distributions on the spin-induced quadrupole moment parameter $\delta\kappa_{s}$, estimated from the observed gravitational wave events GW151226~\cite{GW151226} and GW170608~\cite{GW170608}. Left and right panels correspond to two different physically motivated  priors on $\delta\kappa_{s}$ parameter (symmetric and one-sided). The posteriors are obtained from the Bayesian analysis of the O1/O2 public GW data using \lalinference\cite{Veitch:2014wba}. We used \imrPhenomPv2 waveform models~\cite{Husa:2015iqa} for the analysis, truncated at the inspiral-to-merger transition frequency as the spin-induced deformations are not modelled in the merger and ringdown phases. The vertical dotted lines show the $90\%$ credible bounds (highest density intervals) on $\delta\kappa_{s}$.}
		\label{fig:realeventspos} \end{figure*}

	In this work, we implement and demonstrate the method~\cite{Krishnendu:2017shb} within the framework of Bayesian inference and perform tests of binary black hole nature of the LIGO-Virgo detected binary black hole events. Our method uses binary black hole
	waveforms with parametrized deformations on the spin-induced quadrupole
	moment coefficients $\kappa$,  defined as
	$\kappa=1+\delta\kappa$ where the parametrized deformations (labeled as $\delta\kappa$) represents the deviations from binary black hole nature.  We make use of the \lalinference \cite{LSCAlgorithmLibrarySuite,Veitch:2014wba} library to measure the parameterized deformations $\delta\kappa$ of compact binaries which can be considered as the bounds on their departures from binary black hole natures.    Our method also includes estimation of Bayes factors to perform Bayesian model selection between binary black hole models and black hole mimicker models.      
	
	We perform detailed studies to demonstrate the method using simulated GW signals (injections) which include those of various masses and spins. We investigate in detail about various degeneracies in the parameter space and associated systematics in the estimated parameters, which may often restrict the applicability of this test. Finally, we apply this method on the LIGO-Virgo detected binary black holes  GW151226 and GW170608 and obtain constraints on their BH natures.     
	
	\subsection{Executive Summary: Constraints from GW151226 and GW170608}\label{sec-exec-summary}
	Here we briefly summarize the results from the tests of binary black hole nature of the observed GW signals GW151226~\cite{GW151226}
	and GW170608~\cite{GW170104}. Among  all the ten binary black hole events detected in
	O1/O2, we have restricted the analysis for these two events. This is
	because, with the currently available waveform models, our test is
	applicable only on the inspiral part of the signal and GW151226 and GW170608
	are the only two inspiral dominated events. 
	
	\begin{table}
		\begin{tabular}{lccc}
			\hline\hline
			\addlinespace[1mm]
			\textbf{Event} & \hspace{6mm} \textbf{Prior} \hspace{7mm} & \hspace{7mm}\textbf{$90\%$ bounds}\hspace{4mm} & \textbf{Bayes factor } \\
			& \textbf{on $\dks$ } & \textbf{on $\dks$ } & \textbf{($\log\BgenericVsGR$)}\\
			\hline\hline
			\addlinespace[1mm]
			\multirow{2}{*}{GW151226} & [-200, 200] &[-191.78, 13.45]  &  -0.94 \\
			\addlinespace[1mm]
			& [0, 200] &  $\leq$  98.67 &  -2.26 \\
			\addlinespace[1mm]
			\hline
			\addlinespace[1mm]
			\multirow{2}{*}{GW170608} & [-200, 200] &[-177.36, 122.98]  &  -0.15  \\
			\addlinespace[1mm]
			& [0, 200] &  $\leq$  125.69 &  -1.15 \\        
			\addlinespace[1mm]
			\hline\hline
		\end{tabular}
		\caption{Summary of the tests of binary black hole nature of the real gravitational wave events GW151226 and GW170608 by measuring the spin-induced quadrupole moment parameters $\dks$. The results are shown for two different physically motivated priors on $\dks$: [-200, 200] (symmetric) and [0, 200] (one-sided) as shown in the second column. The third and fourth columns respectively show the $90\%$ credible intervals (upper bounds in case of one-sided priors) on $\delta\kappa_s$ and the Bayes factors between non-BH and BH models.}
		\label{Tab:realevents}
	\end{table}

	Figure \ref{fig:realeventspos} shows the bounds obtained from
	GW151226~\cite{GW151226} (red) and GW170608~\cite{GW170608} (green).
	We show the posterior probability distribution for $\dks$, the parametrized deformations in the $\ks$ parameter, which is the symmetric combination of spin-induced quadrupole moment coefficients of the individual compact objects ($\kappa_1$ and $\kappa_2$), as discussed in section \ref{sec-KAM-summary}. In the left panel, we used a generic
	prior on $\dks$, as uniform in [-200, 200],  which leads to constraints on
	generic BH mimicker models which has positive or negative values for
	$\dks$. Under this prior assumption, we find that the deformation
	parameter $\dks$ is constrained to a 90\% credible interval of
	[ -191.78, 13.45] for GW151226 and [ -177.36, 122.98] for GW170608. In the right panel, we have obtained the bounds on $\dks$ for a restricted one-sided prior of [0, 200]. Unlike the generic prior, this one-sided prior leads to constraints on specific black hole mimicker models such as boson stars for which $\dks$ is predicted to be positive always. Under this prior assumption, we obtain
	90\% credible upper bounds to be $\dks \leq  98.67$ for GW151226 and
	$\dks \leq  125.69$ for GW170608. All the bounds are listed in
	Table~\ref{Tab:realevents}. In all the cases, it is noted that the BH
	limits ($\dks=0$) are well within the 90\% credible intervals which means that the posteriors do not indicate the presence of any non-BH nature in these events. However, one may also note that the posteriors are not very sharply peaked at zero implying weaker constraints on the non-BH nature of the compact objects involved.
	
	In addition to the bounds reported above, we performed Bayesian model selection between BH mimicker models and BH models by calculating the Bayes factor between them (defined in Sec.~\ref{sec:Bayes}). The estimated Bayes factors for both the events are given in Table~\ref{Tab:realevents}.  For these events, we find that the  Bayes factors in the logarithmic scale are -0.94 (for GW151226) and -0.15 (for GW170608) which implies that Bayes factors do not show strong evidence in favor of any of the models (neither BH nor non-BH models). These results are in agreement with our conclusions from the posteriors discussed above. Only more sensitive measurements in the future may help us quantify this better. In section~\ref{sec:GWevents}, we have discussed the results from gravitational wave events in more detail.
	
	The rest of this paper is organized as follows. In
	Section~\ref{sec.3:methodology}, we discuss the waveform model used in this study and give a brief overview of  Bayesian inference for parameter estimation and model selection. Section~\ref{sec:simulations}
	covers our detailed simulation studies and results, and in section
	\ref{sec:GWevents}, we present the constraints obtained from the real events GW151226 and GW170608.
	
	\section{Method} \label{sec.3:methodology}
	
	\subsection{The waveform model} \label{sec-waveform}
	
	In frequency domain, the gravitational wave signal from compact  binary inspirals in the detector frame can be schematically written as,
	\begin{equation}
	\tilde{h}(f) = \mathcal{C}\,\mathcal{A}(f)\, \mbox{\large{\it{e}}}^{i{\mathit{\psi}(f)}},
	\label{eqn:wf}
	\end{equation}
	where, $\mathit\psi(f)$ is the phase and $\mathcal{A}(f)$ is the amplitude of the gravitational wave signal which is given by $\sim
	D_L^{-1}\,\Mc^{5/6}\,f^{-7/6}$     where $\Mc$ is the chirp mass, which is
	related to individual masses $m_{1}$ and $m_{2}$ as,
	$\Mc$=$\frac{(m_{1}\,m_{2})^{3/5}}{(m_{1}+m_{2})^{1/5}}$, and $D_L$ is
	the luminosity distance to the source. The factor $\mathcal{C}$ carries
	the antenna response of the interferometers as a function of the source location and orientation parameters.

	The orbital evolution of the inspiralling binary is largely
	encoded in the phasing formula and appears in terms of the masses and
	spins of the binary {\footnote{ We have not considered the effects due to orbital eccentricity, tidal deformations due to the presence of external gravitational field \etc in the waveform.}}.  Due to the recent developments in the
	post-Newtonian modeling of compact binaries~\cite{Blanchet}, the
	phasing formula for the inspiralling binary has been computed accurately
	up to 3.5PN order
	\cite{Marsat:2012fn,Bohe:2012mr,Bohe:2013cla,Marsat:2013caa,Bohe:2015ana,Marsat:2014xea,Arun:2008kb,Kidder:1995zr,Will:1996zj,Buonanno:2012rv,Blanchet:2013haa,Mishra:2016whh,Buonanno:1998gg,Nagar:2018zoe,EOB_Precessing_2013}.
	
	This phasing formula accounts for the higher-order spin
	corrections such as  spin-orbit interactions (at 1.5PN, 2PN, 3PN and
	3.5PN orders) and spin-spin interactions (at 2PN and 3PN orders).
	Spin-induced quadrupole moment coefficient given in
	Eq.~(\ref{eq-leadingorder}) first appears at the 2PN order and its first
	post-Newtonian 
	correction appears at the 3PN order~\cite{Marsat:2012fn,Bohe:2012mr,Bohe:2013cla,Marsat:2013caa,Bohe:2015ana,Marsat:2014xea}.

	Since the spin-induced quadrupole moment parameter is unity for Kerr BHs, the waveforms which are
	particularly developed for binary black hole systems a priori assume the
	value unity. However, for this study, since our interest is in those
	binary systems for which $\kappa$ departs from \textit{unity}, we
	re-write Eq.~(\ref{eq-leadingorder}) in the following form, 
	\begin{equation}
	Q=-(1+\delta\kappa)\,\chi^2\,m^3, \label{eq--parametrized} 
	\end{equation}
	where $\delta\kappa$ is the parametrized departure of $\kappa$
	from unity. Hence $\delta\kappa=0$ is the BH limit and non-zero
	$\delta\kappa$ corresponds to non-BH objects. Our proposal is to
	independently measure $\delta\kappa$ and use the measurement to put possible constraints on the allowed parameter space of BH mimicker models from observed gravitational wave events. 
	
	For this study, we use the \imrPhenomPv2~\cite{Hannam:2013oca} waveform approximant which is available in LSC Algorithm Library, by incorporating into it, the parametrized deformations shown in Eq.~(\ref{eq--parametrized}). \imrPhenomPv2 is a frequency domain inspiral-merger-ringdown waveform model whose inspiral part of the phasing agrees with the PN phasing and the merger-ringdown parts are obtained by calibrating to the \textit{numerical-relativity} waveforms~\cite{Taracchini:2013rva,Buonanno:2000ef,Hannam:2013oca,Buonanno:1998gg,Ajith:2009bn}. These \textit{numerical-relativity} waveforms have been computed by assuming binary black hole nature (\textit{ie}, $\dks=0$) by default. Therefore the merger and ringdown phases of the \imrPhenomPv2 do not account for the $\kappa$ effects hence the analytical parametrization described in  Eq.~(\ref{eq--parametrized}) is not expected to be valid once the binary enters into the merger regime of the evolution. To avoid any systematic biases due to this, we truncate our analysis at the inspiral-to-merger transition frequency of the \imrPhenomPv2 defined by $f_{\rm{upper}}=0.018/M$, where $M$ is the total mass of the system~\cite{Husa:2015iqa}. As investigated in Ref.~\cite{Ghosh:2017gfp}, we expect  negligible amount of spectral leakage effects due to this sharp cut-off.
	
	\subsection{Choice of test parameters}
	
	In the most general case, each compact object in the binary can have independent spin-induced quadrupole moments $\kappa_1$ and $\kappa_2$ which are different from the Kerr value of unity. Hence we can parametrize a potential deviation of the BH nature
	by introducing two independent deformation parameters
	$\delta\kappa_{1}$ and $\delta\kappa_{2}$ given by $\kappa_{1,2}=1+\delta\kappa_{1,2}$.
	Due to the strong degeneracy between $\delta\kappa_{1,2}$ in the gravitational
	waveform, simultaneous measurement of the two would yield very weak
	constraints~\cite{Krishnendu:2017shb,Krishnendu:2018nqa}. 
	
	Hence one may resort to
	an alternative approach where one of the linear combinations of the
	$\delta\kappa_{1,2}$ parameters is estimated from the data. Following
	\cite{Krishnendu:2017shb}, we consider the symmetric combination
	$\delta\kappa_s=\frac{1}{2}(\delta\kappa_1+\delta\kappa_2)$ as the parameter which captures the deviation from binary black hole nature and estimates the associated error bars when the anti-symmetric combination is zero ($\delta\kappa_1=\delta \kappa_2$).  Though restrictive, this does not weaken the proposed null test because a break down of this assumption is also likely to lead to a shift of the peak of the posterior
	of $\delta\kappa_s$  away from zero which is what we look for as evidence for
	the presence of black hole mimickers.

	\subsection{Overview of Bayesian inference}\label{sec:Bayes}
	We provide a brief review of Bayesian inference for
	gravitational wave parameter estimation and model selection keeping the
	present context of testing the binary black hole nature in mind. The subject has been well described in literature \cite{Veitch:2014wba,JohnVeitch2010bayesian}.
	
	Given the data \textit{d}, which contains the signal and the noise, if $\Hyp$ is our hypothesis (or model) about the signal, then following Bayes' theorem, the posterior probability of the signal parameters $\paramVec$ can be written as, 
	\begin{equation} \label{bayeseqn} 
	P(\paramVec \vert \Hyp, d) = \frac{P(\paramVec \vert \Hyp)\, P(d \vert \Hyp, \paramVec)}{P(d \vert \Hyp)},
	\end{equation}        
	where, $P(\paramVec \vert \Hyp)$ is the {\it{prior probability}} and  $P(d\,\vert \paramVec, \Hyp)$ is the {\it{likelihood function}}.  For a Gaussian wide-sense stationary noise, the {\it likelihood function} can be expressed as, 
	\begin{equation} 
	P(d \vert \Hyp,\,\paramVec) \propto 
	\exp{\left[- \frac{(\td-\th|\td-\th)}{2}\right]}, 
	\end{equation}
	where, $\td$ and $\th$ are respectively the data and model waveform in frequency domain. The noise weighted inner product $(.|.)$ appearing in the exponent is defined as
	\begin{equation} \label{eq-likelihood} 
	(x|y) = 4 \Re
	\int_{\flower}^{\fupper} \frac{x(f)^*y(f)}{\snf} df 
	\end{equation}
	where the $^*$ indicates complex conjugate and $\snf$ is the one-sided power spectral density (PSD) of the noise. The lower limit of integration $\flower$ is the seismic cut-off frequency and the upper limit $\fupper$ is the inspiral-to-merger transition frequency discussed in section \ref{sec-waveform}. $P(d \vert \Hyp )$ in Eq.~(\ref{bayeseqn}) is the {\it{Bayesian evidence}} for the model $\Hyp$, denoted by $\Z$, which is obtained as the likelihood marginalized over the prior volume, 
	\begin{equation}
	\label{eq-evidence} 
	\Z = P(d \vert \Hyp) = \int P(\paramVec \vert \Hyp ) \, P(d \vert \paramVec, \Hyp) d\paramVec, 
	\end{equation} 
	where the integration is over the entire prior volume of the multi-dimensional parameter space. Evidence quantifies how much the data $d$ is in favor of the model $\Hyp$ within the prior domain.  
	
	To test the binary black hole nature of the compact binaries, we define the following two models:
	\begin{enumerate}
		\item 
		The binary black hole model $\Hgr$ which reads as     ``\textit{The source of the gravitational wave signal is binary black holes in general relativity}''. For this model, the waveform assumes $\kappa_s$ to be unity (or  $\delta\kappa_s = 0$) and the set of parameters defining this model (\thatis binary black hole parameters) is denoted as $\thetaBH$. 
		
		\item 
		The non-BH model $\Hgeneric$ which reads as ``\textit{ The source of the gravitational wave signal is a binary of non-BH compact objects \textit{aka} BH mimickers}''. The waveform for this model allows $\kappa_s$ to deviate from unity. Therefore we use $\dks$ as a free parameter and the set of parameters defining this model is given as $\thetaNBH = \{ \thetaBH,\dks\}$. 
	\end{enumerate}
	
	\begin{figure*}[htp] \centering
		\includegraphics[scale=0.58]{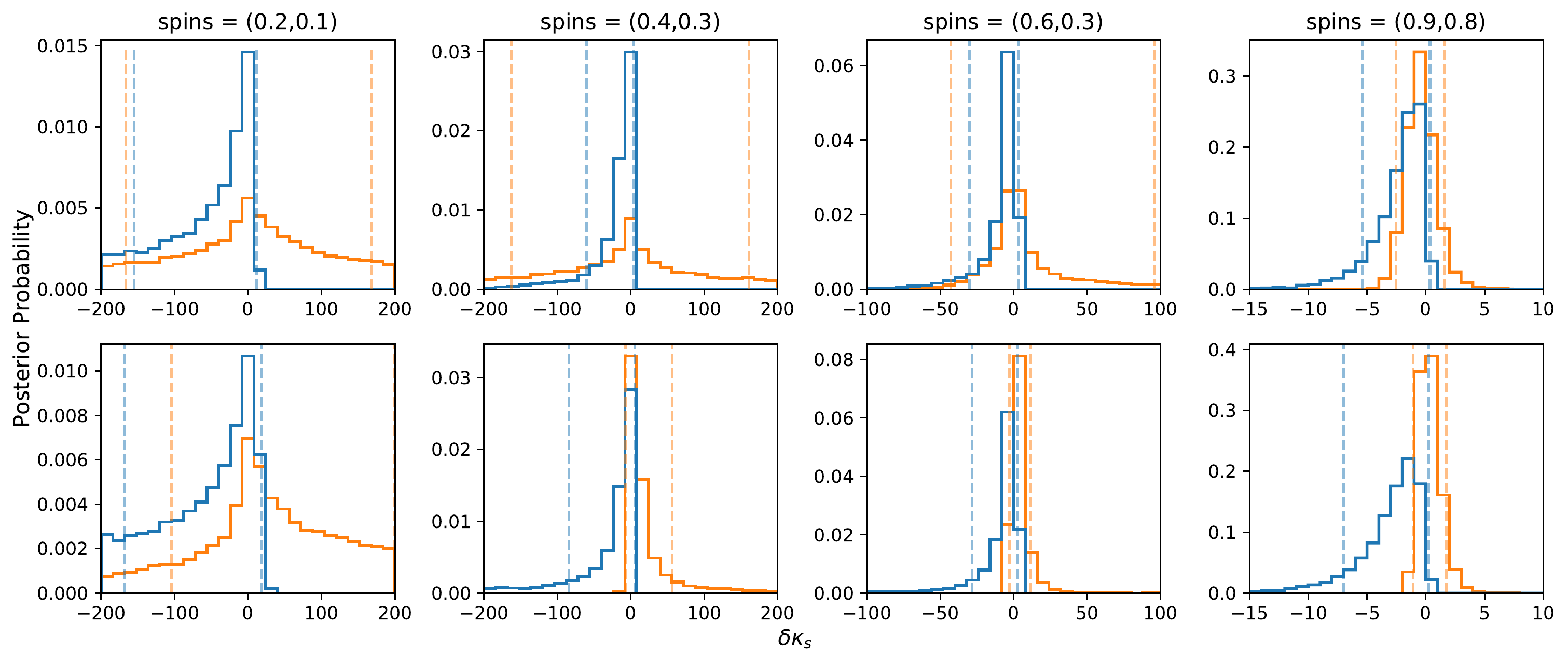} 
		\caption{Posterior distributions on $\dks$ for a binary systems with total mass 15$\Msun$ and mass ratio 1 (top row) and 2 (bottom row) for different spin magnitudes of  (0.2, 0.1), (0.4, 0.3), ( 0.6, 0.3) and ( 0.9, 0.8) from left to right in each row.  Binaries are assumed to be optimally oriented at a luminosity distance of 400 Mpc. Different colours represent different injected spin orientations: both spins aligned to the orbital angular momentum (light blue) and both spins anti-aligned to the orbital angular momentum (orange).}
		\label{figure1manypanels} 
	\end{figure*}    
	
	The one-dimensional posterior for $\delta\kappa_s$ parameter can be obtained by marginalizing the multi-dimensional posterior over the other parameters, \thatis
	\begin{equation} \label{marginalization}
	P(\delta\kappa_s \vert \Hgeneric,\, d ) = \int P\left(\{\thetaBH,\delta\kappa_s\}\vert \Hgeneric,d\right) \, d\thetaBH,
	\end{equation} 
	and the 90\% credible intervals on $\delta\kappa_s$ are obtained as the shortest interval $(\delta\kappa_s^{l},\delta\kappa_s^{r})$ which contains 90\% of the posterior probability distribution, \thatis  
	\begin{equation}\label{eq-bayesian-credible}    
	\int_{\delta\kappa_s^{l}}^{\delta\kappa_s^{r}} P(\delta\kappa_s \,\vert \Hgeneric, \, d)\,  d\delta\kappa_s \sim 0.9.  
	\end{equation}
	
	To perform model selection between the BH and non-BH models, we compute the Bayes factor between $\Hgeneric$ and $\Hgr$ as follows,
	\begin{equation} 
	\BgenericVsGR = \frac{\Zalt}{\Zgr},
	\label{eq-evidence-ratio-2} 
	\end{equation} 
	which quantifies how well the data favors the BH mimicker hypothesis $\Hgeneric$ over the BH hypothesis $\Hgr$. When there is no prior preference for one model over the other, then Bayes factor is same as the odds ratio between the two models (Odds ratio is defined as the ratio of posterior probabilities of the two models \thatis $P(\Hgeneric\vert d)/P(\Hgr \vert d)$). Following definition of evidence in Eq.~(\ref{eq-evidence}), the Bayes factor in Eq.~(\ref{eq-evidence-ratio-2}) can be written as,
	\begin{equation}
	\BgenericVsGR = \frac
	{\int P(\thetaNBH \vert \Hgeneric ) \, P(d \vert \thetaNBH, \Hgeneric) d\thetaBH \, d\dks}
	{\int P(\thetaBH \vert \Hgr ) \, P(d \vert \thetaBH , \Hgr) \, d\thetaBH}.
	\label{eq-BF}
	\end{equation} 
	
	For both parameter estimation as well as model selection
	studies in this paper, we use \lalinference
	\cite{Veitch:2014wba} which is a Bayesian inference package available in
	the LSC Algorithm Library
	\cite{LSCAlgorithmLibrarySuite,Veitch:2014wba}. \lalinference makes use
	of stochastic sampling algorithms such as Nested Sampling
	\cite{skilling2006}, Markov Chain Monte Carlo (MCMC) sampling
	\cite{Feroz:2007kg,Feroz:2008xx,Feroz:2013hea} \etc  and we use Nested Sampling algorithm for the analysis in this study. 
	
	
	\section{Studies using simulated data and results}\label{sec:simulations}
	
	In this section, we perform detailed studies using simulated
	data to assess the efficiency of the proposed method to distinguish
	between binary black holes and binaries comprising of black hole
	mimickers. The aim is primarily to quantify the bounds on the $\delta
	\kappa_s$ parameter as a function of the source parameters of the
	expected gravitational wave signal. We also present the Bayes factors between black
	hole mimickers and black hole models as a function of $\delta \kappa_s$.
	
	\subsection{Details of simulations}
	
	\textbf{Masses:} 
	We choose binary systems with the total mass
	$M=15\Msun$ in the detector frame \footnote{Total mass of $M=15\Msun$ in the detector frame will corresponds to $M\sim13.8\Msun$ in the source frame if we assume the luminosity distance to source to be 400 Mpc.} and mass ratios $q=1$ and $q=2$ as representative cases. The masses are chosen such that they ensure the signals have a significant amount of inspiral in the detector band as the parametrization we employ is in the inspiral part of the waveform.    
	
	\textbf{Spins:} 
	Four combinations of component spins (dimensionless spin magnitudes) are used:  (0.2,  0.1), (0.4, 0.3), (0.6, 0.3) and (0.9, 0.8) which represent low, moderate, and high spins, respectively here the heavier BH in the binary always assumed to be highly spinning compared to the lower mass BH. Each component spin can be either aligned or anti-aligned with respect to the orbital angular momentum vector. Therefore, for each binary we consider four possible spin configurations: both are aligned, the heavier BH spin is aligned but the lighter BH spin is anti-aligned, the heavier BH spin is anti-aligned but the lighter BH spin is aligned and both BH spins are anti-aligned.
	
	\textbf{$\dks$ parameter:} 
	Binary black hole injections are generated with $\dks$=0 while non-BH injections are generated by choosing  $\dks$ in the range [-40, 40]. The non-BH injections are used to compute the Bayes factors between the non-BH and BH hypotheses.
	
	\textbf{Extrinsic parameters:} 
	We choose a fixed distance of  400$\Mpc$ for all the systems which is broadly motivated by the typical distances of several binary black hole mergers during the first two observing runs of Advanced LIGO and Advanced Virgo. 
	For all the systems above, the sky-location and orientation are chosen in such a way that they are optimally oriented and located for the detector network under consideration. Both sky-location and orientation of the source can affect our estimates of $\dks$ only through the signal-to-noise ratio.

	\textbf{Prior choices:} 
	We use prior on  $\delta\kappa_s$ to be uniform in [-200, 200]. This range includes the spin-induced quadrupole moment values predicted for various binary black hole mimicker models~\cite{Ryan97b,Uchikata:2015yma}. Priors on the dimensionless spin parameters (component spins) are chosen such that their magnitudes are uniform in [0, 1] and their directions are isotropically distributed. Component mass priors are uniform in [4, 100]$\Msun$.
	Further, all the injections are non-precessing (\thatis aligned or anti-aligned spins) whereas the recovery waveform models account for precession effects. 
	
	\textbf{Network configuration:} 
	Throughout our studies, we consider a three-detector network (HLV) which includes two advanced LIGO detectors at Hanford (H) and Livingston (L) \cite{AdvancedLIGO,Harry:2010zz,Advanced_LIGO_Reference_Design} and advanced Virgo detector (V) \cite{TheVirgostatus,TheVirgo:2014hva}, assuming both LIGO and Virgo at their design sensitivities given by references \cite{UpdatedAdvancedLIGOsensitivitydesignscurve} and \cite{UpdatedAdvancedVIRGOsensitivitydesignscurve,TheVirgostatus}, respectively.      
	
	\textbf{Zero-noise injections:} 
	Injections are generated using the \texttt{lalsim-inspiral} library available in the LSC Algorithm Library
	\cite{LSCAlgorithmLibrarySuite} with  \imrPhenomPv2 as the waveform approximant. For all the injections, we assume noise realizations to be zero (zero-noise injections) in order to avoid biases in the parameter estimates introduced by a particular noise realization. Results from a zero-noise realization is equivalent to results averaged over many realizations of zero-mean random noise. A noise realization is not to be confused with the noise PSD $S_n(f)$ which appears in the likelihood integral (see Eq.~(\ref{eq-likelihood})) which is always used while computing the relevant quantities.
	
	\textbf{Other details:} 
	For the current analysis, we use a sensitive lower cut-off frequency of $\flower=$20Hz for all three detectors. The upper cut-off frequency $\fupper$ of the integral in Eq. (\ref{eq-likelihood}) is chosen as the inspiral-to-merger transition frequency of the IMRPhenomPv2 waveform which is related to the total mass of the system through the relation $M\,f_{\rm{upper}}=0.018$~\cite{Husa:2015iqa}, as described earlier.   With all these criterions applied, our test injections have the three-detector (LHV) network SNR ranging between $\sim$40-45, where the slight variations are caused by the spins.

 The neglect of the merger and ringdown parts of the
waveform will lead to deterioration in the overall parameter estimation.  As mentioned before, this is why we choose the injections with total mass of  $M=15\Msun$ so that  the SNR contribution from the merger and ringdown is as low as  $\sim 3\%$ of the total SNR, assuming zero spins. For GW151226 and GW170608 on which we have applied the test in this work, the SNR contributions from the merger and ringdown are $\sim 7\%$ and $\sim 5\%$ respectively. 
However, we would like to stress that the neglect of merger and ringdown
in our case is due to the unavailability of a physical model for
spin-induced multipole moments beyond the inspiral. In future, with
better analytical understanding of the dynamics of the remnant black
holes, situation may change (see, for example,  \cite{Gupta:2018znn} for a recent work
along this direction).

	\begin{figure*}[htp] 
		\centering
		\includegraphics[scale=0.5]{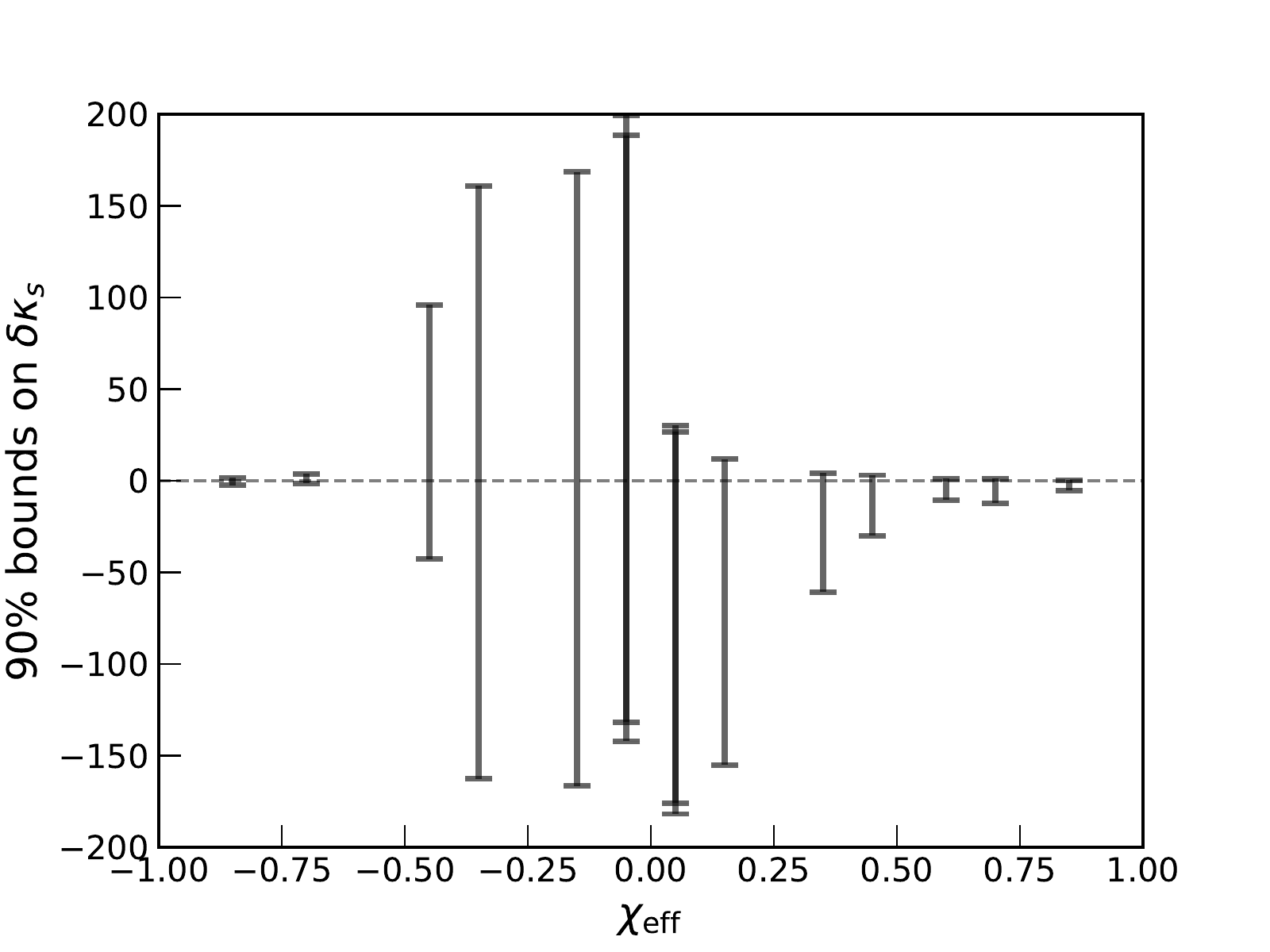}
		\includegraphics[scale=0.5]{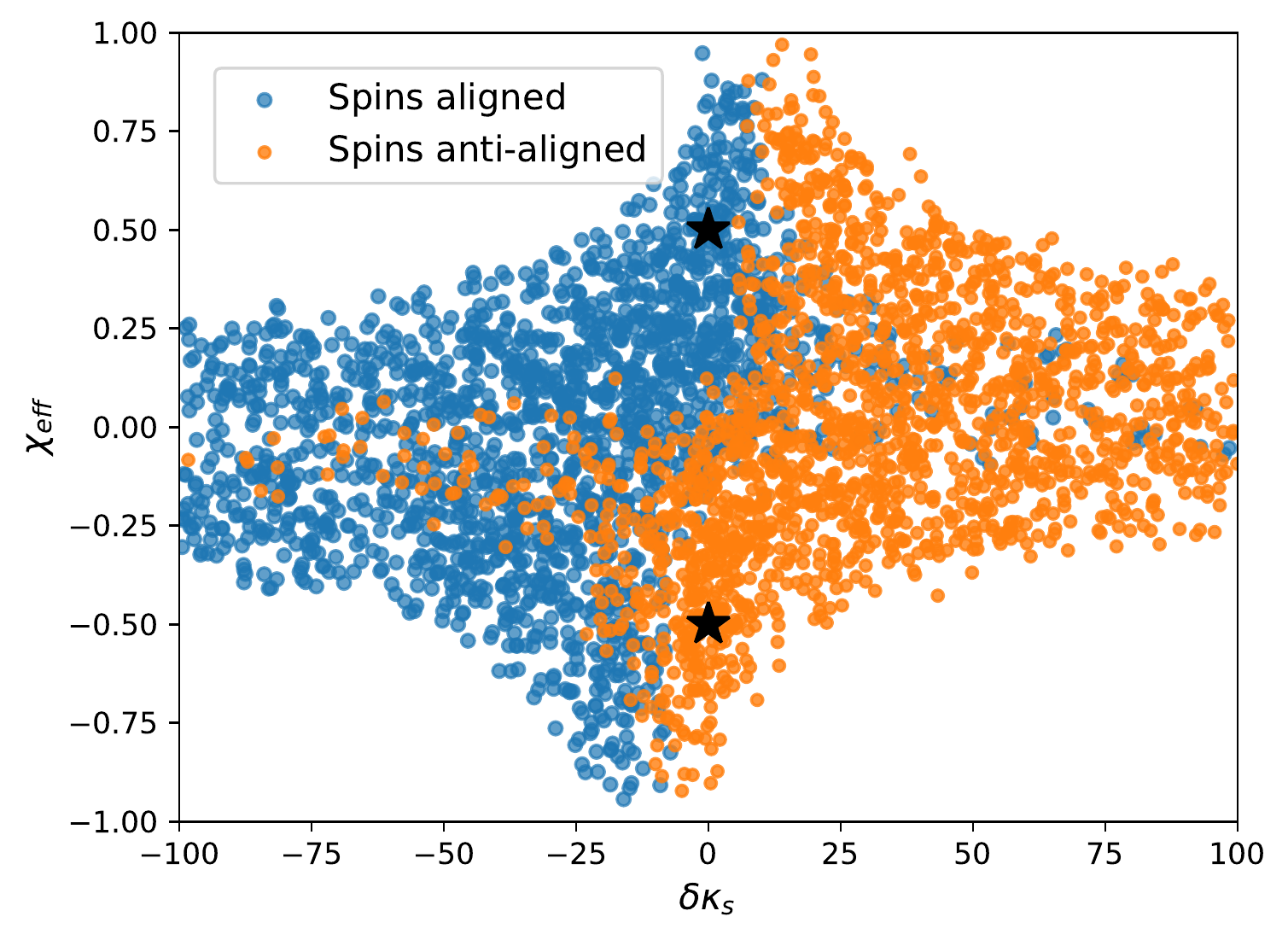}
		\caption{{\bf Left:} The $90\%$ bounds on the spin-induced quadrupole moment parameter $(\dks)$ given in Eq.~(\ref{eq-bayesian-credible}) as a function of the injected valuees of effective spin parameter  (see Eq.~(\ref{effspin})). All the injections are compact binary inspirals with fixed total mass of $15M_{\odot}$ while varying the mass ratio, spin magnitudes and orientations which results in different values of effective spin parameter. 
			{\bf Right:} Figure showing the degenerate regions in the non-BH parameter space ($\dks-\effspin$) for binary black hole injections with two different spin orientations aligned  (0.6,  0.3) and anti-aligned (-0.6, -0.3). The light blue and orange represent aligned and anti-aligned cases respectively and the injected parameters are marked by black stars. The scattered points show the region at which the non-BH waveform has a very high overlap (${\cal O}>0.995$) with the BH injection(s) (See Eq.~(\ref{eq-overlap})).} 
		\label{figure-violin-overlap} 
	\end{figure*}    
	
	\subsection{Bounds on $\dks$ parameter} 
	
	Fig.~\ref{figure1manypanels} shows the posterior probability
	distributions of $\dks$ parameter obtained from the various simulations.
	The first row corresponds to component masses (7.5, 7.5)$\Msun$ (mass
	ratio = 1) and the second row corresponds to component masses (10,
	5)$\Msun$ (mass ratio = 2). In each row, the four different columns
	correspond to four spin magnitudes (0.2,  0.1), ( 0.4,  0.3), (0.6, 0.3)
	and  (0.9, 0.8) from left to right. The different colors represent different injected
	spin orientations: both the spins aligned (light blue) and both spins anti-aligned (orange) to the orbital angular
	momentum axis. The dashed vertical lines are the 90\% credible bounds following the respective colors of the histograms.  Recall that the bounds are estimated
	as the highest density intervals of the posteriors as defined
	in Eq.~(\ref{eq-bayesian-credible}).
	
	It is evident from Fig.~\ref{figure1manypanels} that the bounds on $\dks$ are stronger when the spin magnitudes are larger (see the panels from left to right together with their narrowing axis range). This is expected because, for larger spin magnitudes, the waveform has stronger signatures of spin-induced quadrupole moments (see Eq.~(\ref{eq--parametrized})) which in turn improves the measurement.
	
	Though all the posteriors in Fig.~\ref{figure1manypanels} peak
	at their injected values ($\dks=0$), we notice that there is skewness in all the posteriors about their injected values. This skewness gets mirror-reflected when the spin orientation is reversed. In other words, comparing the light blue and orange histograms in each panel, one notices that the longer tail for light blue is towards left-hand side while for orange, it is towards the right-hand side. This indicates that our ability to constrain the non-BH nature is different for aligned and anti-aligned spin orientations. For aligned cases, the type of non-BH nature with $\dks>0$ (such as binaries of boson stars) can be better constrained than the type of non-BH nature with $\dks<0$ (such as binaries of gravastars). On the other hand, for anti-aligned cases, it is vice versa. We investigate these features in detail below.

	\subsubsection{Role of effective spin parameter}    
	
	We find that the effective spin parameter $\effspin$ plays a major role in the features observed in the posteriors discussed above. Effective spin parameter defined as 
	\begin{equation}
	\chi_{\rm{eff}}=\frac{m_1\,\chi_{1z}+m_2\,\chi_{2z}}{(m_1+m_2)},
	\label{effspin}
	\end{equation}
	is a combination of component masses $m_{1}$, $m_{2}$ and
	component spins $\chi_{1z}$, $\chi_{2z}$ and appears as the leading order
	spin dependence in the inspiral PN waveform~\cite{Ajith:2009bn}. In
	Fig.~\ref{figure-violin-overlap} (left panel), we have shown the bounds
	on $\dks$ parameter as a function of their injected $\effspin$ values where the
	vertical bars correspond to the 90\% credible intervals of the $\dks$
	parameter. The larger the magnitude of $\chi_{\rm eff}$, the tighter the
	bounds on $\dks$. For systems with small magnitudes of $\chi_{\rm eff}$ (for example, $\chi_{\rm eff}<0.3$), the $\dks$ parameter is almost unconstrained. Further, when $\effspin$ is large and positive, the region with $\dks>0$ is better constrained, whereas when the $\effspin$ is large and negative, the region with $\dks<0$ is better constrained.
	
	The dependence of $\dks$ posteriors on $\effspin$ discussed
	above holds true despite the fact that the systems considered for this
	plot include those with various component masses and spins. In fact, it
	is difficult to disentangle the individual effects of the component
	masses and spins due to the degeneracy between spins and mass ratio
	parameters~\cite{Baird:2012cu}. However, $\chi_{\rm eff}$ captures the combined effects of all these parameters on the $\dks$ posteriors and hence is the most important single parameter which describes our ability to constrain $\dks$ parameter for any given system.  
	
	We further investigate the skewness of the posteriors in detail and show that they are primarily caused by the waveform degeneracies between $\dks$ and $\effspin$ parameters. To demonstrate this, we first define the overlap function $\mathcal{O}$ between a binary black hole injection $\tilde{h}^{\rm{BH}}$ and a non-BH template $\tilde{h}^{\rm{NBH}}$ as, 
	\begin{equation}
	\mathcal{O} = \frac{\left( \hBH | \hNBH \right)}
	{\sqrt{ \left( \hBH | \hBH \right)
			\left( \hNBH|\hNBH \right)  }}
	\label{eq-overlap}
	\end{equation}
	where $(.|.)$ is the noise weighted inner product defined in Eq.~(\ref{eq-likelihood}) and both $\hBH$ and $\hNBH$ are in frequency domain. Overlap quantifies how similar are the two signals $\hBH$ and $\hNBH$ and its value is maximum ($\mathcal{O}=1$) when $\hBH = \hNBH$.

	\begin{figure*}[htp] \centering
		\includegraphics[scale=0.5]{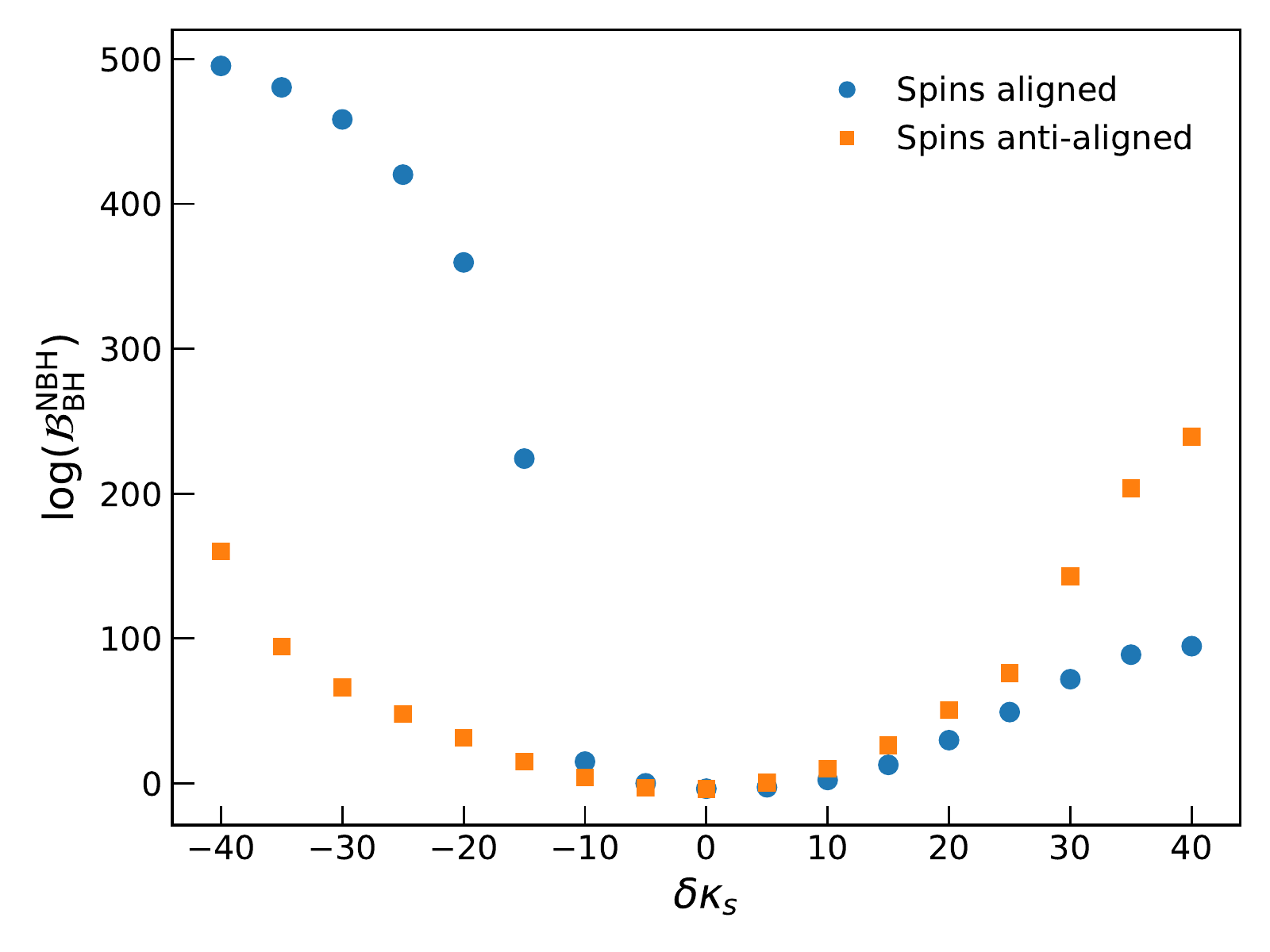}
		\includegraphics[scale=0.5]{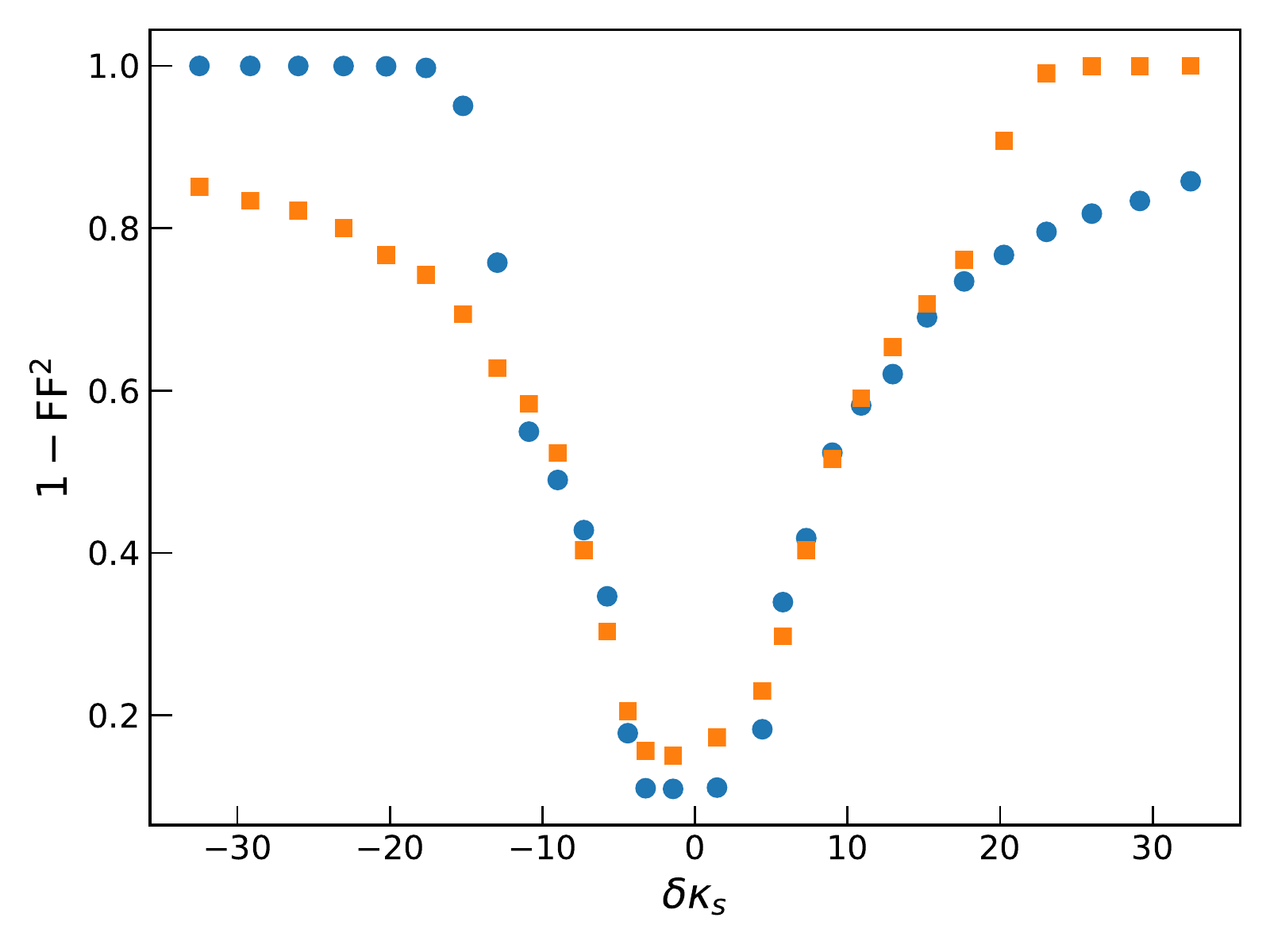}
		\caption{{\bf Left: } Demonstration of Bayes factor between non-BH and BH models for different non-BH injections. The x-axis shows the injected value of $\dks$ and y-axis shows the log of Bayes factors. All the injections are of component masses $(10+5)M_{\odot}$ and fixed spin magnitudes $(0.6, 0.3)$ while the light blue and orange markers correspond to aligned and anti-aligned  spin orientations respectively. {\bf Right: } 
			Complementary analysis done using \textit{fitting factors} motivated by \cite{Vallisneri:2012qq}. For each non-BH injection with values of $\dks$ as given on x-axis, the \textit{fitting factor} $\ff$ was computed \wrt the BH waveforms by maximizing the overlap over the BH parameter space. Here the maximization is done on a restricted BH parameter space with $\effspin$ being the only free parameter, with the remaining parameters fixed to their injected values. The quantity on y-axis is $1-\ff^2$ which explicitly appears in the approximate scheme of \cite{Vallisneri:2012qq} (See Eq.~(\ref{eq-BF-FF}))} 
		
		\label{fig-BF-FF} 
	\end{figure*}
	
	We have taken two binary black hole injections with both of them having identical component masses $(10, 5)M_{\odot}$ but different spin orientations (0.6, 0.3) and (-0.6, -0.3) whose  $\effspin$ values are $0.5$ and $-0.5$ respectively. The templates $\tilde{h}^{\rm{NBH}}$ are uniformly distributed in the non-BH parameter space with component spins ranging in [-1, 1] and $\dks$ ranging between [-100, 100]. The masses of the templates are kept fixed at their injection values which will be justified later with the results. 
	
	We show the results of this overlap calculation in the right
	panel of Fig.~\ref{figure-violin-overlap}. Templates having very high
	overlaps with the injections ($\mathcal{O}>0.995$) are shown as
	scattered plots in the $\dks-\effspin$ plane (light blue for aligned-spin
	injection and orange for anti-aligned spin injection). The injected
	parameters are marked with stars (black color). For the aligned spin case (light blue), there are more scattered points on the left half ($\dks<0$) compared to the right half ($\dks>0$). This indicates that the left half is more degenerate and hence less distinguishable from the injected binary black hole signal, compared to the right half. This is exactly the feature observed in the posteriors as well as the bar plot (Fig.~\ref{figure1manypanels} and Fig.~\ref{figure-violin-overlap} left) that for systems with aligned spins (or $\effspin>0$), the positive side of the $\dks$ posterior is better constrained than the negative side. A similar explanation holds for the anti-aligned spin case as well with all the features turned exactly opposite. 
	
	The fact that the masses of the templates are fixed to the injected values might be considered as ignoring some of the other potential degeneracies which are present. However, ignoring the role of such degeneracies can be justified since we have shown above that the $\dks-\effspin$ degeneracies could solely explain the features of the posteriors. In other words, the overlap study with masses fixed to the injections helps underline that it is the $\dks-\effspin$ degeneracy which is primarily responsible for the features of the posteriors.

	\subsection{Model selection between BH and non-BH models} 
	\label{sec.4:Detectability}    
	
	In this section, we discuss the model selection studies between non-BH and BH models by obtaining Bayes factors between them. We estimate the Bayes factors $\BnonGRvsGR$ (see Eq.~(\ref{eq-BF})) using \lalinference for a set of non-BH injections whose $\dks$ varies in the range [-40, 40]. All the injections are of fixed component masses (10, 5)$\Msun$ while the analysis is repeated with two spin choices for the injections: (0.6, 0.3) and (-0.6, -0.3), which as followed in the previous section, represents the aligned and anti-aligned orientations respectively. 
	
	The results are shown in the left panel of Fig.~\ref{fig-BF-FF} where the log of the Bayes factors ($\log\BnonGRvsGR$) are plotted as a function of the injected $\dks$ values. The light blue and orange colors correspond to aligned and anti-aligned spins respectively. As one would expect, when the magnitude of the injected $\dks$ increases, the Bayes factor increases which mean that they can be better distinguished from binary black hole models. We notice that the way Bayes factor increases with $\dks$ is different for aligned and anti-aligned cases. For example, among all the injections with $\dks>0$ (such as binaries of boson stars), Bayes factors are larger for those whose spins are anti-aligned (or negative $\effspin$) compared to those whose spins are aligned (or positive $\effspin$). The reverse is true for the injections with $\dks<0$ (such as binaries of gravastars). 
	
	The features discussed above can have possible consequences on the identification of BH mimicker populations. For example, among the population of boson star binaries, our ability to distinguish them from binary black holes will be inclined towards those with anti-aligned spins. As a result, the population which we identify as binary boson stars will have more sources with anti-aligned spins (or negative $\effspin$). Similarly, the population which we identify as binary gravastars will have more sources with aligned spins (or positive $\effspin$).

	\subsubsection{Further investigations using Fitting Factor}

	In order to investigate various features in the Bayes factor
	plot (Fig.~\ref{fig-BF-FF}, left panel), we perform a study using
	\textit{fitting factor} to complement the Bayesian analysis. The \textit{fitting factor} of a non-BH waveform $\hNBH$, with a BH waveform model $\hBH$ is given by,
	\begin{equation}
	\ffNBH = \max_{\thetaBH}\left( \frac{\left( \hNBH | \hBH(\thetaBH) \right)}
	{\sqrt{ \left( \hNBH | \hNBH \right)
			\left( \hBH(\thetaBH) | \hBH(\thetaBH) \right)  }}\right)
	\label{eq-FF}
	\end{equation}
	where $\hNBH$ is evaluated at a given point $\thetaNBH$ in the
	non-BH parameter space and $\thetaBH$ is any arbitrary point in the BH
	parameter space over which the maximisation is carried out. One can see
	from Eq.~(\ref{eq-FF}) that $\ffNBH$ is equal to the \textit{overlap},
	defined in Eq.~(\ref{eq-overlap}), maximised over the BH parameter
	space. Qualitatively, $\ffNBH$ is regarded as a measure of how well the
	BH waveform model $\hBH$ can mimic the given non-BH signal
	$\hNBH(\thetaNBH)$. In other words, $\ffNBH$ describes how well the
	non-BH corrections contained in $\hNBH(\thetaNBH)$ can be
	re-absorbed\footnote{We closely follow the terminology used by
		Vallisneri~\cite{Vallisneri:2012qq} here.} into the BH waveform $\hBH(\thetaBH)$, by varying $\thetaBH$ within its allowed range.

	As discussed before, the Bayes factor $\BnonGRvsGR$ for a given
	signal is high when the signal has a non-BH component of the form that
	can not be re-absorbed into the BH waveform. Broadly this implies
	that a high Bayes factor is closely related to a low \textit{fitting factor}. Cornish {\it et al}.~\cite{Cornish:2011ys} and Vallisneri {\it et al}.~\cite{Vallisneri:2012qq} showed an approximate scheme to relate the Bayes factor and \textit{fitting factor} which, for our context (considering only the dominant term in the expression) would read as,   
	\begin{equation}
	\log \BnonGRvsGR \propto \rho^2 \times  \left( 1-\ffNBH^2 \right)
	\label{eq-BF-FF}    
	\end{equation}
	where $\rho$ is the signal-to-noise ratio. In a later work, Del Pozzo et al.~\cite{DelPozzo2014} explored this in more detail using numerical simulations and extended its validity regimes by introducing additional correction terms. 
	
	In this exercise, we consider a set of non-BH injections similar
	to the ones considered in the Bayes factor studies above. For all the
	injections, we compute $\ffNBH$ using Eq.~(\ref{eq-FF}) for a binary
	system of masses (10, 5)\msun. Note that in our case the BH parameter
	space ($\thetaBH$) over which the maximization is done has only one free
	parameter which is the effective spin $\effspin$, while all other
	parameters are fixed to their injected values as our goal is to understand the ability of $\effspin$ to mimic non-BH signals.
	
	The results are shown in the right panel of Fig.~\ref{fig-BF-FF}
	where  $1-\ffNBH^2$  (which explicitly appears in Eq.~i(\ref{eq-BF-FF}))
	is plotted as a function of the injected $\dks$. We find similar
	features as seen in the Bayes factor plot (left panel). For example, for
	non-BH injections with $\dks>0$, the value of $(1-\ff^2)$ is higher for
	anti-aligned cases (or negative $\effspin$ cases) while it is the
	opposite for those injections which have $\dks<0$. That means the
	results independently obtained from the Bayes factor and the \textit{fitting factor} analyses are complementary to each other. We emphasize
	again that this agreement holds despite restricting the BH parameter
	space to just one parameter, $\effspin$.

	Thus, the \textit{fitting factor} analysis further underscores the key
	role played by the $\effspin$-$\dks$ degeneracy in distinguishing non-BH binaries from BH binaries. 
	
	Notice that when the non-BH signals are mimicked by
	the BH waveforms, it happens at the cost of offsets in the estimated  BH
	parameters from their true values. In realistic cases, this will result
	in systematic biases in the estimated BH parameters, if BH waveforms are
	used for the analysis while the true signal was of a non-BH binary. It
	is worth mentioning the two contexts in which this can happen. 1) if one
	presumes the signal to be of BH nature and hence ignore the possibility
	of any potential non-BH nature. 2) one does not assume BH nature a
	priori, however, given the SNR of the signal, the non-BH component in the signal is mild enough to be
	reabsorbed into the BH waveform by varying the parameters. Though both
	the biases are fundamental in nature~\cite{Yunes:2009ke,Cornish:2011ys},
	the former is also the result of our prior assumption while the latter is the result of our waveform models being insufficient to account for the underlying non-BH effects or/and the non-BH effects being buried in noise. In a follow-up work, these effects will
	be investigated in detail.  
	
	\begin{figure*}[htp] \centering
		\subfigure{\includegraphics[scale=0.36]{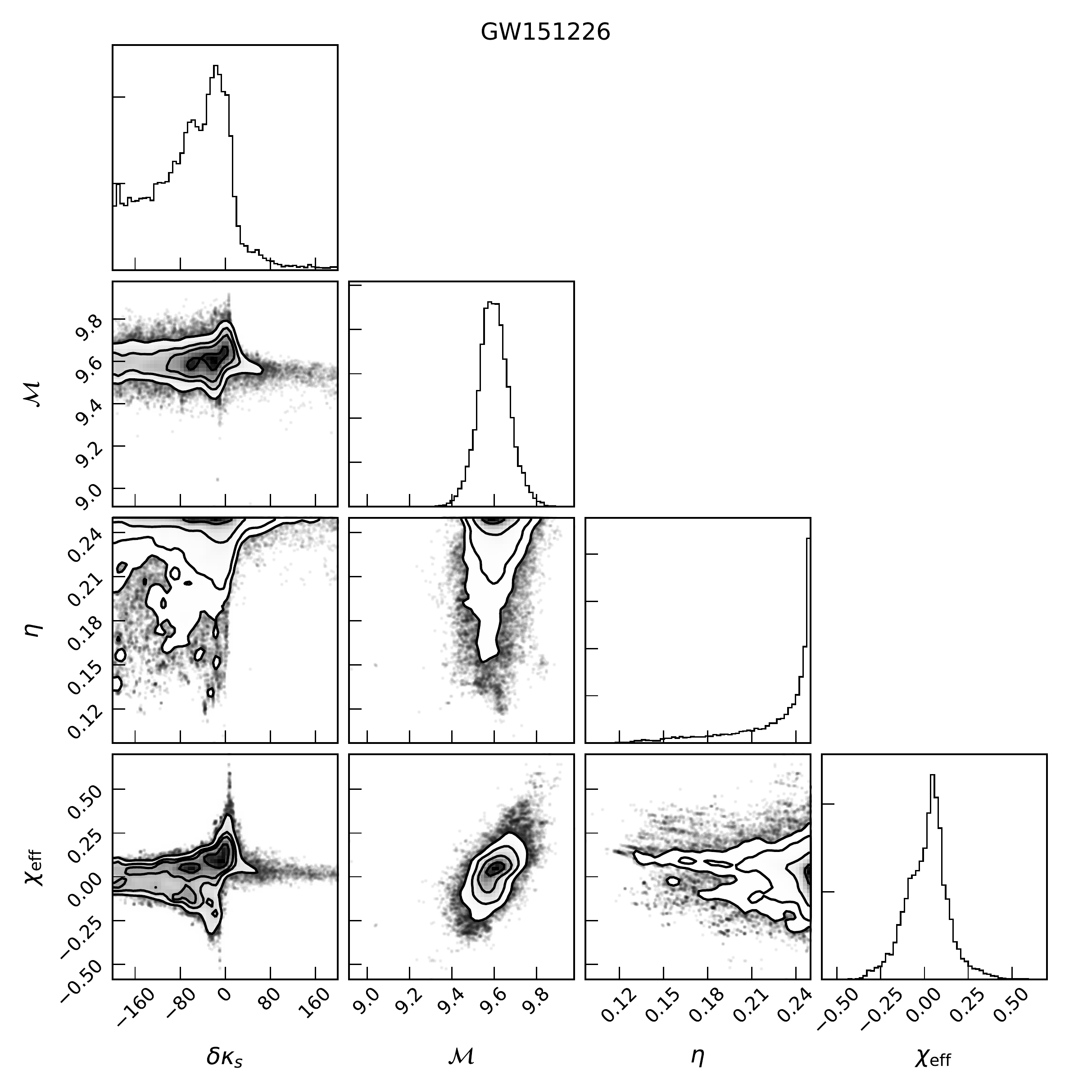}}
		\subfigure{\includegraphics[scale=0.36]{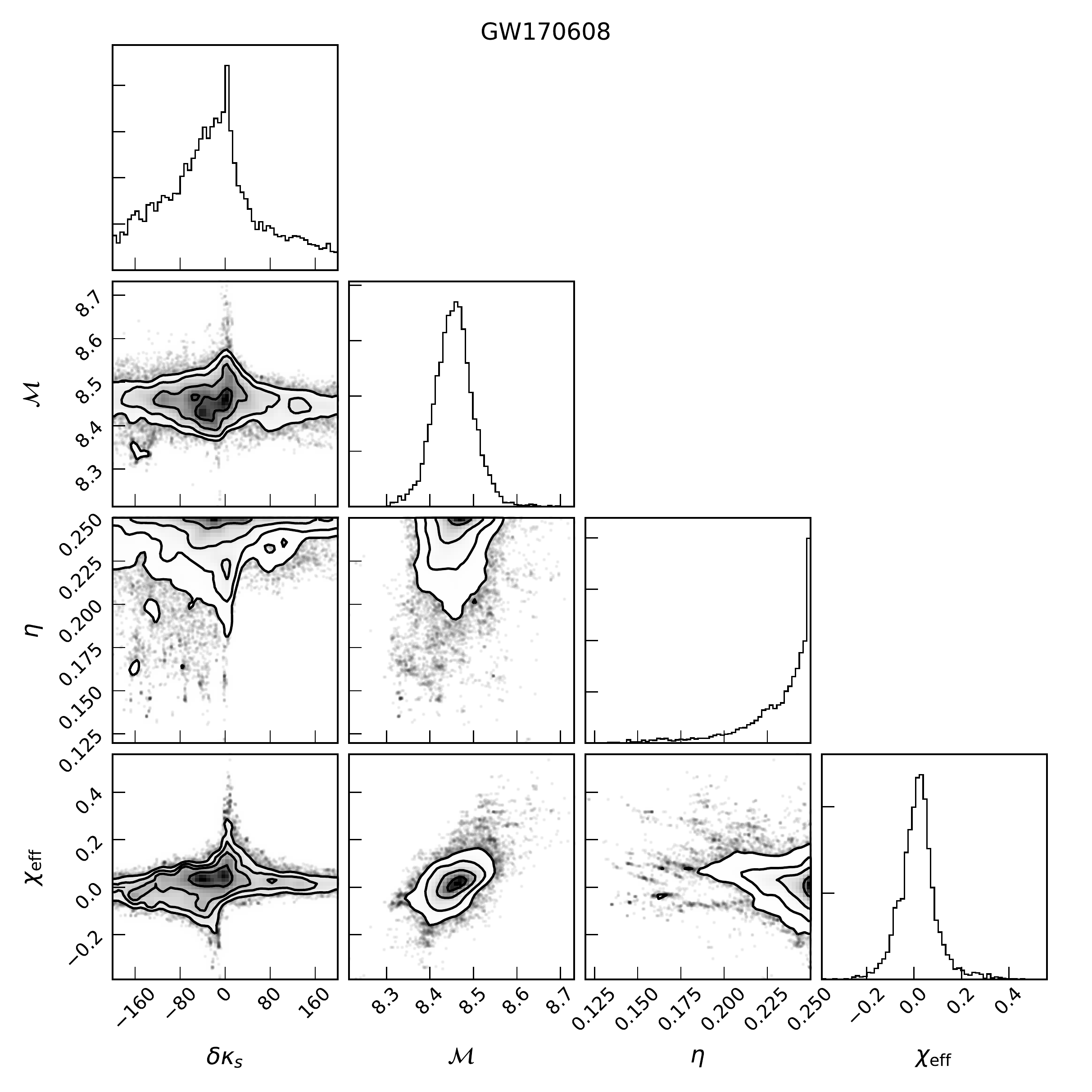}}
		\caption{The corner plots of $\delta\kappa_s$, chirp mass ($\Mc$), symmetric mass ratio ($\eta$) and $\chi_{\rm{eff}}$ from GW151226~\cite{GW151226} and GW170608~\cite{GW170608} with symmetric priors on $\delta\kappa_s$. 
		} 
		\label{fig:realevent_cornerplots} 
	\end{figure*}

	\section{Testing the binary black hole nature of GW151226 and GW170608}\label{sec:GWevents}
	
	As reported in \cite{O1O2catalogLSC2018}, the first two
	observation runs (O1/O2) of Advanced LIGO and Advanced Virgo have
	identified ten gravitational wave signals which are consistent with
	binary black hole waveforms. In this section, we apply the proposed
	spin-induced quadrupole moment test on some of the observed events
	and
	ask how consistent they are to the binary black hole hypothesis. As
	discussed before, at present, our test is based on a parametrization of the inspiral part of the waveform. Therefore, we restrict the study to the two inspiral-dominated signals GW151226~\cite{GW151226} and GW170608~\cite{GW170608}, where the inspiral only signal to noise ratio is at least $\sim 10$. The estimated (median) detector frame total mass of GW151226 and GW170608 are $23.55 \Msun$ and $19.89 \Msun$~\cite{GWOSC} and the corresponding inspiral-to-merger transition frequencies are $155.12$Hz and $184.55$Hz respectively. We use these as the upper cut-off frequencies ($\fupper$) of the analyses along with  \imrPhenomPv2 as the waveform approximant.
	
	The main results are shown in Fig.~\ref{fig:realeventspos} where
	the posteriors on $\dks$ parameter are discussed (as we 
	summarized in section~\ref{sec-exec-summary}). We consider two different
	priors on $\dks$: a symmetric prior $[-200, 200]$ (left panel) and a
	one-sided prior $[0, 200]$ (right panel). The symmetric prior [-200, 200]
	represents a most generic test which accounts for BH mimicker models
	including those of both oblate ($\dks>0$) and prolate ($\dks<0$)
	spin-induced deformations. The one-sided prior [0, 200] is a restricted
	case which accounts only for oblate spin-induced deformations. In other
	words, the symmetric prior leads to generic constraints on BH mimicker
	models including boson stars, gravastars \textit{etc.} whereas the
	one-sided prior is  motivated by specific models such as boson star models for which $\dks$ is always positive and hence meant to provide specific constraints on such models. The prior is restricted to $\vert \delta\kappa_{s}\vert\leq200$ because the parametrized waveforms we construct are found not to be well-behaved beyond this range and hence cannot meaningfully represent the corresponding physics. The 90\% credible intervals (highest density intervals) on $\dks$ are given in Table~\ref{Tab:realevents}. For all the cases, it is found that the 90\% credible intervals or the upper bounds (in case of one-sided prior) are consistent with $\dks$ being equal to zero and hence consistent with GW151226 and GW170608 being binary black holes. 
	
	Detailed corner plots are presented in Fig.~\ref{fig:realevent_cornerplots} which will help us to gain further insights about the underlying degeneracies and correlations. As we discussed earlier, the $\dks$ parameter is found to be highly degenerate with $\effspin$. Again, we note that the posteriors of $\dks$ are asymmetric about their most probable values and both the events have got more posterior support for negative values of $\dks$ than positive values. We recall from Fig.~\ref{figure1manypanels} and \ref{figure-violin-overlap} that the similar posterior features were observed for cases in which positive values of $\effspin$ were injected. As seen in the corner plots, the estimated (median) $\effspin$ values are positive for both these events and hence the results from these two events are completely consistent with our findings from simulation studies.     It is also found that the $\dks$ posteriors are railing against the prior boundaries for both the events.  This may improve in future if there are events which have larger spins or lower masses, similar to the ones considered in the simulations earlier.
	
	We also performed Bayes factor studies on both the events whose
	results are also shown in Table~\ref{Tab:realevents}. With the symmetric
	and the one-sided priors on $\dks$, we computed Bayes factors
	($\BnonGRvsGR$) between the non-BH and BH models ($\Hgeneric$ and $\Hgr$
	respectively). We find that the log of the Bayes factors
	($\log\BnonGRvsGR$) are in the range $-2.3 < \log\BnonGRvsGR < 0$ for all
	the cases. These values are too small to be considered as evidence for
	favoring or rejecting any of the models which are tested. The slightly
	negative values obtained in all the cases may be interpreted as weak
	evidence in favor of BH models over non-BH models. We notice that these
	features are consistent with those observed in the posteriors in
	Fig.~\ref{fig:realeventspos} that the posteriors are spread over a wider
	range of values of $\dks$ with significant weights over non-BH
	(\textit{i.e.} non-zero) values. 
	
	\section{Conclusions and Outlook}
	In this work, we have developed a Bayesian framework to test the binary
	black hole nature of gravitational wave signals using the measurements
	of spin-induced quadrupole moment parameters of the compact binaries as
	proposed in Ref.~\cite{Krishnendu:2017shb}. We carried out detailed
	studies using simulated gravitational wave signals to test the
	applicability of our method. The waveform models which are used for this
	test currently includes spin-induced deformation terms only in the
	inspiral part and hence its applicability is limited to the inspiral
	regime. 
	
	We applied the method on the two inspiral-dominated events from O1/O2,
	GW151226 and GW170608, and obtained bounds on their binary blackhole
	natures. These are the first constraints on the black hole nature of the
	compact binaries detected by Advanced LIGO and Advanced Virgo. With more
	gravitational wave detections with inspiral-dominated signals,
	especially of higher spins, there will be increased opportunity to
	perform the tests of BH nature using spin-induced quadrupole moment parameter measurements. 

 It may require the sensitivity of the third-generation detectors
such as ET or CE to put very tight constraints on possible
deviations from BBH nature of the compact binaries. These detectors are
likely to achieve bounds on $\kappa_s$ of the order of unity for spins
as low as 0.2~\cite{Krishnendu:2018nqa} which should be sufficient to
rule out a large class of BH mimickers which predict $\kappa_s$ to be
${\cal O}(10)$. 
Since NSs may have $\kappa\sim 10$, this method may  also be useful in distinguishing NS-BH binaries from binary black holes. Future work will address this topic and the issue of analyzing systems where the results depend on the mass ratio and the spin of the BH in addition to the $\kappa$ of the NS. 
However, there may still be some space left for objects
like gravastars~\cite{Uchikata2016qku} which predict $\kappa_s$ to be
${\cal O}(1)$. Space-based interferometers like LISA or DECIGO may be
able to set limits on $\kappa_s$ as stringent as  ${\cal O} (0.1)$ for moderate
spins~\cite{KY19,Krishnendu:2017shb} which may be sufficient to rule out most of the
known black hole mimicker models.
 
 However, some more work will have to be done before these bounds
can be used to constrain the parameter space of BH mimicker models such
as boson stars. Firstly, the limit on the maximum spin for BHs does not apply to
boson stars which can have dimensionless spin magnitudes greater than 1
(see for example ~\cite{Ryan97b}). Secondly, boson stars can have
quartic dependence on spins of the form, say, $Q=-
m^3\chi^2(\kappa+\epsilon\,\chi^4)$. These aspects may have to be
incorporated into the analysis for deriving meaningful
bounds on the allowed parameter space of boson stars. There are ongoing
efforts to address these issues.

	\acknowledgements 
	
	We are extremely grateful to P. Ajith and  N. J-McDaniel for their useful
	comments and suggestions at every stage of the analysis.  We
	thank A. Gupta and  N. J-McDaniel for carefully reading the manuscript and giving comments.
	N. V. K. acknowledges
	useful discussion with A. Ghosh, Sumit Kumar, and M. K. Haris. 
	N.V.K would like to thank International Center for Theoretical Sciences,
	Bangalore for hospitality during the initial stages of the project.
	The authors acknowledge the use of LIGO LDG clusters for the computational work done for this study.
	K. G. A.,  N. V. K, M. Saleem were partially supported by a
	grant from Infosys Foundation. K. G. A also acknowledges the
	Swarnajayanti fellowship grant 
	DST/SJF/PSA-01/2017-18 and 
	EMR/2016/005594 of SERB.  K. G. A. acknowledges the Indo-US Science and
	Technology Forum through the Indo-US {\em Centre for the Exploration of Extreme
		Gravity}, grant IUSSTF/JC-029/2016.   AS is supported by the research program of the
	Netherlands Organisation for Scientific Research (NWO).   This document has
	LIGO preprint number {\tt P1900215}. 

\appendix

\bibliography{ref}
\bibliographystyle{apsrev}

\end{document}